\documentclass[prc,unsortedaddress,superscriptaddress,amsmath,amsfonts,amstex,amssymb,showpacs,floatfix]{revtex4}
\usepackage{graphics,epsfig}
\begin{document}

\title{Once more about astrophysical $S$ factor for the $\alpha + d \to {}^{6}{\rm Li} + \gamma$ reaction.}

\author{A.\,M.~Mukhamedzhanov}
\affiliation{Cyclotron Institute, Texas A\&M University,
College Station, TX 77843}

\author{B. F. Irgaziev}
\affiliation{2GIK Institute of Engineering Sciences and Technology, Topi, Pakistan}

\begin{abstract}
Recently to study the radiative capture $\alpha + d \to {}^{6}{\rm Li} + \gamma$  process a new measurement of the ${}^{6}{\rm Li}({\rm A\,150\,MeV})$ dissociation in the field of ${}^{208}{\rm Pb}$ 
 has been reported in [F. Hammache {\it et al.} Phys. Rev ${\bf C 82}$, 065803 (2010)]. However, the dominance of the nuclear breakup over the Coulomb one prevented from obtaining the information about  the $\alpha + d \to {}^{6}{\rm Li} + \gamma$  process from the breakup data. The astrophysical $S_{24}(E)$ factor has been calculated within the $\alpha-d$ two-body potential model with potentials determined from the fits to the $\alpha-d$ elastic scattering phase shifts. However, the scattering phase shift itself doesn't provide a unique $\alpha-d$ bound state potential, which is the most crucial input when calculating the $S_{24}(E)$ astrophysical factor at astrophysical energies.  In this work we emphasize an important role of the asymptotic normalization coefficient (ANC) for ${}^{6}{\rm Li} \to \alpha + d$,
which controls the overall normalization of the peripheral $\alpha + d \to {}^{6}{\rm Li} + \gamma$ process and is determined by the adopted $\alpha-d$ bound state potential.
We demonstrate that the ANC previously determined from the $\alpha-d$ elastic scattering $s$-wave phase shift in [Blokhintsev {\it et. al} Phys. Rev. {\bf C 48}, 2390 (1993)]  gives $S_{24}(E)$, which is at low energies about $38\%$ lower than the one reported in [F. Hammache {\it et al.} Phys. Rev ${\bf C 82}$, 065803 (2010)]. We recalculate also the reaction rates, which are also lower than those obtained in [F. Hammache {\it et al.} Phys. Rev ${\bf C 82}$, 065803 (2010)].  
\end{abstract}


\maketitle

The radiative capture $\alpha + d \to {}^{6}{\rm Li} + \gamma$ is the only process that produces 
${}^{6}{\rm Li}$ in the big bang model. A special interest to this reaction has been trigerred by almost three order disagreemnet between the observational ratio ${}^{6}{\rm Li}/{}^{7}{\rm Li}$
\cite{asplund} and the calculated one \cite{asplund}.  Unfortunately direct measurements of the 
$\alpha + d \to {}^{6}{\rm Li} + \gamma$  radiative capture are practically impossible at astrophysically relevant $\alpha-d$ relative kinetic energies $E \leq 300$ keV due to extremely low cross section.  
Hence, only indirect approach could be feasible to get information about 
${}^{6}{\rm Li}$ formation. The first indirect information about the astrophysical factor 
$S_{24}(E)$ for the    
$\alpha + d \to {}^{6}{\rm Li} + \gamma$ process has been obtained in \cite{kiener} from the Coulomb breakup process ${}^{6}{\rm Pb}({}^{6}{\rm Li\,(26 A\,MeV}),\,\alpha\,d){}^{208}{\rm Pb}$. However, the energy behavior of the extracted astrophysical factor at low energies turned out to be constant what contradicted to all the calculations showing significant drop \cite{muk95}. Recently in \cite{hammache} a new attempt has been done to get the astrophysical factor $S_{24}(E)$ at astrophysically relevant energies from ${}^{6}{\rm Pb}({}^{6}{\rm Li\,(150 A\,MeV}),\,\alpha\,d){}^{208}{\rm Pb}$.
However, analysis has shown significant dominance of the nuclear breakup over the Coulomb one making impossible to determine the needed information about $S_{24}(E)$. Nevertheless, in \cite{hammache}
the $S_{24}(E)$ has been calculated using the $\alpha-d$ two-body potential model. The potentials which are required to make such calculations were obtained from fitting the $\alpha-d$ elastic scattering phase shift for the $s,\,p$ and $d$ waves. The approach used to calculate the astrophysical factor is not related with the studied ${}^{6}{\rm Li}$ breakup process. The only common information in the analysis of the breakup data and calculation of the astrophysical factor were the same bound state and scattering $\alpha-d$ potentials used to generate the corresponding bound state and scattering wave functions. In the potential approach used in \cite{hammache} the bound state potential, as we will discuss below, is the most crucial part of the input, which affects the overall normalization of the astrophysical factor. Unfortunately, the dominance of the nuclear breakup and dependence of the breakup data analysis on the optical potentials doesn't allow one to make a test of the quality of the adopted $\alpha-d$ potential.     
The approach applied in \cite{hammache} is just repetition of the procedure used in \cite{mohr}.  

Here we would like to discuss how reliable are the astrophysical factor $S_{24}(E)$ and the reaction rates for the $\alpha + d \to {}^{6}{\rm Li} + \gamma$  radiative capture reported in \cite{hammache} and what should be done to improve our knowledge about them. It has been long ago recognized that the
$\alpha + d \to {}^{6}{\rm Li} + \gamma$ process at astrophysical energies is entirely peripheral reaction in the two-body potential model \cite{muk95}. Evidently, the potential model itself is a limitation and it would be nice to check peripherality of this reaction within a many-body ab initio approach similat to what has been done recently for the ${}^{3}{\rm He} + {}^{4}{\rm He} \to {}^{7}{\rm Be} + \gamma$ in \cite{neff}. However, an important issue in a many-body approach remains to be solved is reproduction of the experimental binding energy, because the calculated astrophysical factor is sensitive to its value.  Since such ab initio many-body calculations are not yet available we have to live with a more simple 
two-body potential model. The matrix element for the $\alpha + d \to {}^{6}{\rm Li} + \gamma$ direct radiative capture in the long-wave approximation is given by \cite{muk95}
\begin{align} 
M^{\lambda}_{l_{f}\,l_{i} J_{i}} = A_{\lambda}\,\int_{R_{c}}^{\infty}\,dr I_{\alpha\,d\,(01)}^{{}^{6}{\rm Li}}\,r^{\lambda+2}\,\psi_{l_{i}\,J_{i}}(r).
\label{matrelement1}
\end{align}
All the notations are given in \cite{muk95}. The cut-off radius $R_{c}$ is introduced to reflect a peripheral character of the process. The multipolarity of the transition is $\lambda=1,2$. It has been shown in \cite{muk95} that the matrix element shows a very small sensitivity for $R_{c} \leq 4.5$ fm, i.e. until distances which exceeds the ${}^{6}{\rm Li}$ radius and makes possible to approximate 
the radial overlap function by the $C_{\alpha d(01)}\,W_{-\eta,\,1/2}(2\,\kappa_{\alpha d}\,r)/r$, where $C_{\alpha d (01)}$ is the asymptotic normalization coefficient (ANC) for the virtual decay ${}^{6}{\rm Li} \to \alpha + d$ in the state with relative orbital angular momentum $l=0$ and the total angular momentum $J=1$, $W_{-\eta,\,1/2}(2\,\kappa_{\alpha d}\,r)$ is the Whittaker function determining the radial shape of the overlap function beyond of the $\alpha-d$ nuclear interaction region, $\eta$ is the Coulomb 
$\alpha-d$ bound state parameter, $\,\kappa_{\alpha\,d}= \sqrt{2\,\mu_{a\,d}\,\varepsilon_{\alpha\,d}}\,$ is the $\alpha-d$ bound state wave number, $\,\mu_{\alpha\,d}$ and $\,\varepsilon_{\alpha\,d}$ are the reduced mass and binding energy of $\alpha$ and $d$, correspondingly. As we can see, there are two inputs needed to calculate the matrix element. One is the $\alpha-d$ potential describing the continuum in the partial waves $l=1,2$. This potential has been found in cite \cite{hammache} by fitting the elastic scattering phase shifts in these partial waves and it is a legitimate procedure. Note that at very low energies, say around the most effective energy of $70$ keV, one can use a pure Coulomb scattering wave function in the initial state of the reaction. However, the role of nuclear interaction becomes very important with energy increase and it is responsible for reproduction of the resonance at $E_{R}=0.7117$ MeV. In the potential approach used in \cite{hammache,mohr} the overlap function is replaced by the 
\begin{align}
I_{\alpha\,d\,(01)}^{{}^{6}{\rm Li}}= S_{\alpha\,d\,(01)}^{1/2}\,\varphi_{\alpha\,d(101)}(r),
\label{singlpart1}
\end{align}
Here $S_{\alpha\,d\,(01)}$ is the spectroscopic factor for the configuration $(\alpha\,d)_{01}$ in ${}^{6}{\rm Li}$, $\,\,\varphi_{\alpha\,d(101)}(r)$ is the radial wave function of the $\alpha-d$ relative motion in the field generated by the Woods-Saxon potential. Also $n=2$ is the principal quantum number, i.e. the number of the nodes for $r>0$ is $N=1$,  $\,l=0$ and $J=1$ . The spectroscopic factor reflects the fact that the overlap function is not an eigenfunction of any Hamiltonian and, hence, is not normalized to unity in contrast to the bound state wave function. The potential, which is used to calculate the bound state wave function, has been determined in \cite{hammache} from the fitting to the $\alpha-d$ elastic scattering phase shift in the channel $l=0,\,J=1$.
Since the experimental elastic scattering phase shift includes the many-body effects of the scattered nuclei, the same is true for the two-body potential which fits the elastic scattering data. Hence, the spectroscopic factor in Eq. (\ref{singlpart1}) should be set to $S_{\alpha\,d\,(01)}=1$. It is exactly what has been done in \cite{hammache}. Since the reaction under consideration is peripheral at astrophysical energies $E \leq 300$ keV, functions in Eq. (\ref{singlpart1}) can be replaced by their tales, i.e. 
\begin{align}
I_{\alpha\,d\,(01)}^{{}^{6}{\rm Li}} \stackrel{r > R_{c}}{\approx} C_{\alpha d(01)}\,\frac{W_{-\eta,\,1/2}(2\,\kappa_{\alpha d}\,r)}{r}= b_{\alpha d(101)}\,\frac{W_{-\eta,\,1/2}(2\,\kappa_{\alpha d}\,r)}{r},
\label{asympt1}
\end{align}
i.e. in the two-body potential model in the asymptotic region $C_{\alpha d(01)}= b_{\alpha d(101)}$, where $b_{\alpha d(101)}$ is the single-particle ANC for the $\alpha-d$ bound state with the number of nodes for $r > 0$ $\,\,N=1$. The value of the single-particle ANC depends on the adopted bound-state potential. From the parameters of the bound state potential given in \cite{hammache} we find that $b=2.7$ fm${}^{-1/2}$. Thus from the Woods-Saxon potential given in \cite{hammache} we get the ANC, which is about $17\%$ larger then the ANC used in \cite{muk95}. Since the astrophysical factor is proportional to the square of the ANC the usage of $C_{\alpha d(01)}^{2}=7.29$ fm${}^{-1}$ rather than $C_{\alpha d(01)}^{2}=5.29$ fm${}^{-1}$ leads to the increase of the astrophysical factor compared to the one obtained in \cite{muk95} by almost $38\%$.
It can be seen from Fig. 9 \cite{hammache}, where both astrophysical factors are shown in the logarithmic scale. We came to the main point of this paper, the ANC, which is the most crucial input in the calculation of the $S_{24}(E)$. In \cite{hammache}, as in \cite{mohr}, the $s$-wave scattering phase shift has been used to determine the bound-state potential, which generates the bound state wave function and, correspondingly, the ANC. However, it is well known from the inverse scattering problem 
that there is infinite number of the phase-equivalent potentials and to single out a unique potential one has to add two parameters (if only one bound state is present in the given partial wave): the binding energy and the ANC. In \cite{hammache} the adopted potential reproduces experimental $\alpha-d$ binding energy but still, one parameter, the ANC is still missing. Hence, the potential found in \cite{hammache} is one out of an infinite set of the phase equivalent potentials and the ANC, which it generates, is not necessarily a correct one. However, there is another way of using the elastic scattering data which has been realized in \cite{blokh93}. This approach is based on the analiticity of the elastic scattering $S$-matrix element what allows one to extrapolation the experimental scattering amplitude to the bound state pole in the momentum plane to get the residue, which is expressed in terms
of the ANC. Once the ANC has been determined a unique potential can be found \cite{blokh93,blokh2008}. This potential has to satisfy the condition 
\begin{equation}
\mathop {\lim }\limits_{r \to \infty }\,V^{N}(r)\,{e^{2\kappa r}} \to 0,
\label{potentialcond1}
\end{equation}
where $\kappa$ is the bound state wave number. These procedures had been realized in 
\cite{blokh93} for the $\alpha-d$ $\,s$-wave elastic scattering. First, the Pade approximation was used to interpolate the elastic scattering amplitude in the physical region, which was extrapolated to the bound state pole to get the ANC. The obtained ANC  
$C_{\alpha d(01)}= 2.3 \pm 0.12$ fm${}^{-1/2}$ is lower than $C_{\alpha d(01)}= 2.7$ fm${}^{-1/2}$ following from the potential used in \cite{hammache}. After that the two-body potential has been found, which reproduces the $s$-wave  $\alpha-d$ phase shift and provides correct binding energy and the same ANC. 
The found potential is more complicated than a Woods-Saxon one. It is a complex, energy independent potential, which can be expanded in terms of the harmonic oscillator basis. Evidently, that such a potential satisfies condition (\ref{potentialcond1}). Thus the ambiguity in the two-body potential 
calls for more thorough selection of the potential because eventually the adopted potential for the bound state will determine the overall normalization of the astrophysical factor for peripheral direct radiative capture processes.  

\begin{figure}
\epsfig{file=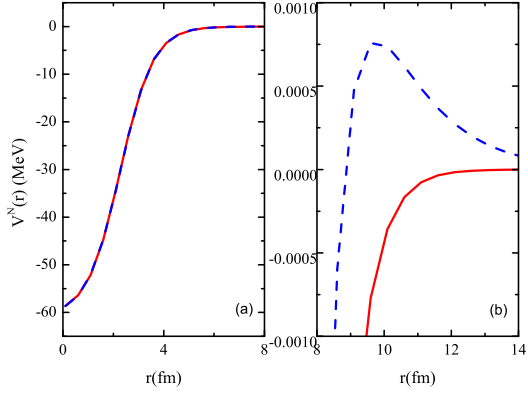,width=8cm}
\caption{(Color online) The red solid line is $\alpha -d$ bound state nuclear potential $V^{N}(r)$ \cite{hammache} and the blue dashed line is $V_{1}^{N}(r)$. The explanation see in the text.}
\label{fig_VN}
\end{figure}

If we would confine ourselves to astrophysical energies then it would be enough to replace in the matrix element (\ref{matrelement1}) the overlap function by its asymptotic
term (\ref{asympt1}) and, evidently, the astrophysical factor will be proportional to 
$C_{\alpha d(01)}^{2}$. However, if we want to extend our calculations to higher energies including the resonance region and above we need to carry more accurate calculations. To compare our results with the $S_{24}(E)$ factor presented in \cite{hammache} we adopt the same $\alpha-d$ scattering potential as in \cite{hammache}. 
For the $\alpha-d$ bound state potential it would be logically reasonable to use the bound state potential from \cite{blokh93,kukulin90}. However, since it has quite complicated form,
we simply modify the bound state potential used in \cite{hammache} using the theorem of the inverse problem in scattering theory \cite{shadansabatier}. This theorem allows one to recover a phase equivalent potential to the Woods-Saxon potential used in \cite{hammache} with arbitrary ANC. Assume that we find a nuclear potential $V^{N}(r)$, which together with the added Coulomb potential $V^{C}(r)$ fits the elastic scattering phase shift in the $l=0$ partial wave and the bound state wave function calculated with this potential has the ANC $C$. Then the phase-equivalent potential is given by
\begin{align}
V_{1}^{N}(r)=  V^{N}(r) + \frac{{\rm d}^{2}K(r)}{{\rm d}r^{2}},
\label{vn11}
\end{align}
\begin{align}
K(r)=   {\rm Log}[1 + (\tau - 1)(1- \int_{0}^{r}\,{\rm d}x\,x^{2}\,\varphi^{2}(x))].
\label{kr1}
\end{align}
The bound state wave function in the potential $V_{1}(r)= V_{1}^{N}(r) + V^{C}(r)$ can be expressed in terms of the bound state wave function $\varphi(r)$  in the potential $V(r)= V^{N}(r) + V^{C}(r)$:
\begin{align}
\varphi_{1}(r)= \tau^{1/2}\,\frac{\varphi(r)}{1 + (\tau -1)\, \int_{0}^{r}\,{\rm d}x\,x^{2}\,\varphi^{2}(x)}.
\label{varphi11}
\end{align}
From Eq. (\ref{varphi11}) taking the limit $r \to \infty$ we obtain that $C_{1}=\tau^{-1/2}\,C$. Let $V(r)=V^{N}(r) + V_{C}(r)$ be the nuclear Woods-Saxon plus Coulomb bound-state $\alpha-d$ potential adopted in \cite{hammache}, which generates the bound-state wave function with $C_{\alpha d(01)}=2.7$ fm${}^{-1/2}$. Then for $\tau =1.378$ we obtain the wave function in the potential $V_{1}(r)$, which has the same ANC as obtained in \cite{blokh93} and used in \cite{muk95}. In Fig. \ref{fig_VN} both nuclear potentials are shown. Since the difference between the potentials is very small, panel (a), in panel (b)
we show the tails of both potentials in the scale allowing to see the difference. 
Note that the existence of the potential $V_{1}^{N}(r)$, which is the phase equivalent to the bound state potential found in \cite{blokh93,kukulin90} with the same ANC, doesn't contradict 
to the inverse scattering theorem because the addition $\frac{{\rm d}^{2}K(r)}{{\rm d}r^{2}}$
to the potential $V^{N}(r)$ asymptotically decays as $\exp(2\,\kappa_{\alpha\,d}\,r)$ violating condition (\ref{potentialcond1}). It doesn't allow one to use potential $V_{1}^{N}(r)$
for analytical extrapolation of the scattering amplitude generated by this potential to the bound state pole but we need this potential here only to generate the bound state wave function
with correct amplitude of the tail, i.e. the ANC.

\begin{figure}
\epsfig{file=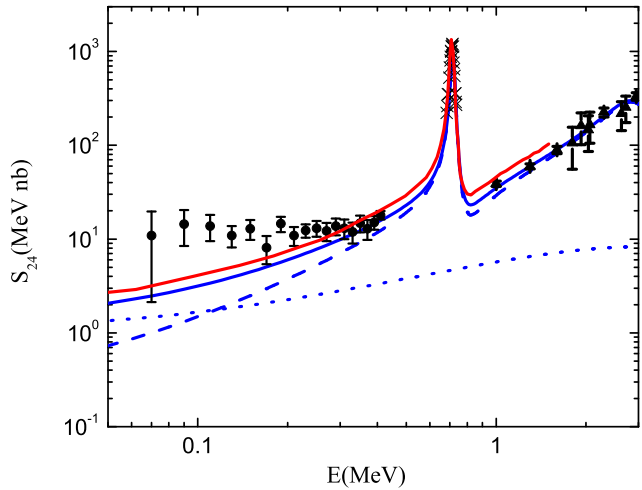,width=10cm}
\caption{(Color online) The astrophysical factors $S_{24}(E)$ for the radiative capture $\alpha + d \to {}^{6}{\rm Li} + \gamma$. Black dots are data from \cite{kiener}; black crosses are data from \cite{mohr}; black triangles are data from \cite{robertson}. The red solid line is the $S_{24}(E)$ factor from \cite{hammache}. The blue dotted line is $S_{24}(E1)$ factor, the blue dashed line is $S_{24}(E2)$ and the blue solid line is our total astrophysical factor $S_{24}(E)$.}
\label{fig_Sfactor}
\end{figure}

Using the bound state wave functions generated by the potentials $V^{N}(r)$ and $V_{1}^{N}(r)$ and the scattering $\alpha-d$ wave functions in the partial waves $l_{i}=1,2$, 
we calculated the $S_{24}(E)$ for the radiative capture $\alpha + d \to {}^{6}{\rm Li} + \gamma$. The results of the calculations are shown in Fig. \ref{fig_Sfactor}.
The astrophysical factor with the bound state wave function generated by the potential $V(r)$
is the $S_{24}(E)$ presented in \cite{hammache} while our $S_{24}(E)$ is the one obtained using the bound state wave function generated by $V_{1}(r)$. At energies below the resonance practically the tails of the bound states wave functions do contribute to the matrix element. Since the square of the ANC in \cite{hammache} is higher than our one by $38\%$, correspondingly the astrophysical factor from \cite{hammache} is systematically higher than our $S_{24}(E)$. Our calculations, definitely better reproduce the experimental data \cite{robertson} at energies larger than the resonance energies, where the calculations from \cite{hammache}
clearly overestimate the data. At resonance energies both calculations reproduce the data \cite{mohr} very well.  At astrophysically relevant energies $E \leq 300$ keV our $S_{24}(E)$ is lower than the one from \cite{hammache} by $38\%$. 
Finaly in Table \ref{table_rates} we present the $\alpha + d \to {}^{6}{\rm Li} + \gamma$ reaction rates, which are also systematically lower than those presented in \cite{hammache}. 
\begin{table}
\caption{The rate for $\alpha + d \to {}^{6}{\rm Li} + \gamma$ reaction
calculated using our astrophysical S-factor for the temperature
range $10^{6}\rm{K}\leq\rm{T}\leq 10^{10}\rm{K}$.\label{table_rates}}.
\begin{ruledtabular}
\begin{tabular}{cccc}
\,\,\,\,$T_9$ \,\,\,\,& $N_a\langle\sigma v\rangle$ & \,\,\,\,$T_9$ \,\,\,\, & $N_a\langle\sigma v\rangle$  \\
 &($cm^3\,mol^{-1}\,s^{-1}$) &  &($cm^3\,mol^{-1}\,s^{-1}$) \\\hline
 0.001& $6.467\times 10^{-30}$& 0.260& $6.823\times 10^{-04}$\\
0.002& $1.857\times 10^{-23}$& 0.270& $7.876\times 10^{-04}$\\
0.003& $2.470\times 10^{-20}$& 0.280& $9.032\times 10^{-04}$\\
0.004& $2.286\times 10^{-18}$& 0.290& $1.030\times 10^{-03}$\\
0.005& $5.693\times 10^{-17}$& 0.300& $1.167\times 10^{-03}$\\
0.006& $6.592\times 10^{-16}$& 0.310& $1.317\times 10^{-03}$\\
0.007& $4.651\times 10^{-15}$& 0.320& $1.478\times 10^{-03}$\\
0.008& $2.327\times 10^{-14}$& 0.330& $1.652\times 10^{-03}$\\
0.009& $9.067\times 10^{-14}$& 0.340& $1.840\times 10^{-03}$\\
0.010& $2.923\times 10^{-13}$& 0.350& $2.040\times 10^{-03}$\\
0.011& $8.127\times 10^{-13}$& 0.360& $2.254\times 10^{-03}$\\
0.012& $2.008\times 10^{-12}$& 0.370& $2.482\times 10^{-03}$\\
0.013& $4.508\times 10^{-12}$& 0.380& $2.725\times 10^{-03}$\\
0.014& $9.343\times 10^{-12}$& 0.390& $2.983\times 10^{-03}$\\
0.015& $1.811\times 10^{-11}$& 0.400& $3.256\times 10^{-03}$\\
0.016& $3.318\times 10^{-11}$& 0.500& $6.930\times 10^{-03}$\\
0.017& $5.787\times 10^{-11}$& 0.600& $1.271\times 10^{-02}$\\
0.018& $9.676\times 10^{-11}$& 0.700& $2.148\times 10^{-02}$\\
0.019& $1.559\times 10^{-10}$& 0.800& $3.462\times 10^{-02}$\\
0.020& $2.432\times 10^{-10}$& 0.900& $5.385\times 10^{-02}$\\
0.025& $1.538\times 10^{-09}$& 1.000& $8.079\times 10^{-02}$\\
0.030& $6.277\times 10^{-09}$& 1.100& $1.166\times 10^{-01}$\\
0.035& $1.929\times 10^{-08}$& 1.200& $1.618\times 10^{-01}$\\
0.040& $4.870\times 10^{-08}$& 1.300& $2.164\times 10^{-01}$\\
0.045& $1.066\times 10^{-07}$& 1.400& $2.797\times 10^{-01}$\\
0.050& $2.093\times 10^{-07}$& 1.500& $3.508\times 10^{-01}$\\
0.060& $6.375\times 10^{-07}$& 1.600& $4.288\times 10^{-01}$\\
0.070& $1.554\times 10^{-06}$& 1.700& $5.126\times 10^{-01}$\\
0.080& $3.245\times 10^{-06}$& 1.800& $6.002\times 10^{-01}$\\
0.090& $6.057\times 10^{-06}$& 1.900& $6.915\times 10^{-01}$\\
0.100& $1.038\times 10^{-05}$& 2.000& $7.854\times 10^{-01}$\\
0.110& $1.665\times 10^{-05}$& 2.100& $8.808\times 10^{-01}$\\
0.120& $2.533\times 10^{-05}$& 2.200& $9.773\times 10^{-01}$\\
0.130& $3.690\times 10^{-05}$& 2.300& $1.074\times 10^{+00}$\\
0.140& $5.185\times 10^{-05}$& 2.400& $1.171\times 10^{+00}$\\
0.150& $7.071\times 10^{-05}$& 2.500& $1.268\times 10^{+00}$\\
0.160& $9.398\times 10^{-05}$& 3.000& $1.745\times 10^{+00}$\\
0.170& $1.222\times 10^{-04}$& 3.500& $2.210\times 10^{+00}$\\
0.180& $1.559\times 10^{-04}$& 4.000& $2.673\times 10^{+00}$\\
0.190& $1.954\times 10^{-04}$& 4.500& $3.145\times 10^{+00}$\\
0.200& $2.416\times 10^{-04}$& 5.000& $3.631\times 10^{+00}$\\
0.210& $2.946\times 10^{-04}$& 6.000& $4.645\times 10^{+00}$\\
0.220& $3.552\times 10^{-04}$& 7.000& $5.689\times 10^{+00}$\\
0.230& $4.238\times 10^{-04}$& 8.000& $6.725\times 10^{+00}$\\
0.240& $5.008\times 10^{-04}$& 9.000& $7.723\times 10^{+00}$\\
0.250& $5.868\times 10^{-04}$& 10.00& $8.664\times 10^{+00}$\\
\end{tabular}
\end{ruledtabular}
\end{table}

In this work we have demonstrated that a crucial quantity, which is necessary to pinpoint the $S_{24}(E)$ astrophysical factor, is the ANC for the virtual decay ${}^{6}{\rm Li} \to \alpha +d$. Due to the peripheral character of the $\alpha + d \to {}^{6}{\rm Li} + \gamma$ direct radiative capture, this ANC determines the overall normalizatiion of the astrophysical factor at astrophysically relevant energies. From our calculations and Fig. \ref{fig_Sfactor} we can see that at low energies the contribution from the isospin forbidden $E1$ transition dominates over the allowed $E2$ transition.
For example, at $E=70$ keV, which is the most effective energy, the contribution from the $E1$ transition to the total $S_{24}(E)$ astrophysical factor is about $60\%$. Even at $E=100$ keV the $E1$ transition contributes about $52\%$ to the total astrophysical factor. Meantime, even if the Coulomb breakup of ${}^{6}{\rm Li}$ would dominate, at $E=70$ keV the $E1$ transition will be suppressed compared to the $E2$ by a factor of $60$. It can hardly make possible to determine the total astrophysical factor from the ${}^{6}{\rm Li}$ experiment.  Since the ANC is the only crucial information needed to calculate the $S_{24}(E)$  astrophysical factor at astrophysical energies, we call for more accurate measurements of the $s$-wave $\alpha-d$ elastic scattering phase shift at lower energies. It will help to extrapolate more accurately the data to the bound state pole to get more ANC for ${}^{6}{\rm Li} \to \alpha +d$.  
Finally, we note that the problem of determination of the two-body bound state potential from the elastic scattering phase shift is quite important in different applications of nuclear reaction theory, in particular, in Faddeev approach for reactions with composite particles.

\section{Acknowledgments}  
The work was supported by the US Department of Energy under Grants No. DE-FG02-93ER40773 and DE-SC0004958 (topical collaboration TORUS) and NSF under Grant No. PHY-0852653.

\end{document}